# PROFILE CONTROL OF A BOROSILICATE-GLASS GROOVE FORMED BY DEEP REACTIVE ION ETCHING

*Teruhisa Akashi and Yasuhiro Yoshimura*

Mechanical Engineering Research Laboratory (MERL), Hitachi, Ltd.

## ABSTRACT

Deep reactive ion etching (DRIE) of borosilicate glass and profile control of an etched groove are reported. DRIE was carried out using an anodically bonded silicon wafer as an etching mask. We controlled the groove profile, namely improving its sidewall angle, by removing excessively thick polymer film produced by carbon-fluoride etching gases during DRIE. Two fabrication processes were experimentally compared for effective removal of the film: DRIE with the addition of argon to the etching gases and a novel combined process in which DRIE and subsequent ultrasonic cleaning in DI water were alternately carried out. Both processes improved the sidewall angle, and it reached $85^o$ independent of the mask-opening width. The results showed the processes can remove excessive polymer film on sidewalls. Accordingly, the processes are an effective way to control the groove profile of borosilicate glass.

## 1. INTRODUCTION

Borosilicate glass, e.g., Coring 7740 Pyrex®, Hoya SD-2, and Asahi Techno Glass SW-3, is a common material for MEMS devices, and it is typically used for a fluidic device and wafer-level packaging (WLP) of an inertial sensor. The WLP usually has a borosilicate-glass cap wafer with cavities and through-holes formed by ultrasonic drilling or sandblasting. These technologies do not enable fabrication that is as precise as etching.

Li et al. first presented their research on deep reactive ion etching (DRIE) of Pyrex® [1], but not many studies have successfully used anisotropic deep dry etching of borosilicate glass, in contrast with silicon [2-4] and silica [5,6]. The reason for this difference is largely due to the quite low selectivity of an etching mask used for borosilicate glass. The selectivity of the mask, e.g., poly-silicon, tungsten silicide, and chromium, is around 20. This means that the necessary thickness of the mask film is calculated as 15 μm when a 300-μm-deep groove is formed. The stress has to be strictly controlled to successfully deposit a 15-μm-thick mask film. Furthermore, the mask needs to be precisely etched. These are the reasons that a borosilicate-glass cap wafer with a deep cavity and a through-hole fabricated by anisotropic etching has not been used for WLP.

The aforementioned low selectivity of an etching mask for DRIE of borosilicate glass limits the groove depth. A novel fabrication process that uses an anodically bonded silicon wafer as an etching mask reportedly overcomes the low selectivity and achieves much deeper etching [7]. In the process, carbon-fluoride gases, i.e., $C_4F_8$ and $CHF_3$, were used as etching gases. However, the drawbacks of this process were that the sidewall angle of the etched groove did not reach more than $80^o$ and that it depended on the mask-opening width. This was due to thick polymer film produced by carbon-fluoride plasma during DRIE. The film was thickly deposited on sidewalls and protected them against the plasma. DRIE needs to remove excessive polymer film to control the etched profile and to improve the sidewall angle.

This paper reports on two fabrication processes for controlling the etching profile, namely for effectively removing excessive polymer film. The obtained experimental results are compared with previous ones [7].

## 2. ETCHING EQUIPMENT AND CONDITIONS

Dry-etching equipment, which is NE500 and is commercially made by ULVAC Inc., was used for the experiment. A schematic of the equipment is shown in Figure 1, and the DRIE conditions are listed in Table 1. The equipment consists of a setting room where a 4-inch wafer is placed and an etching chamber where inductively coupled plasma is generated. The wafer for etching was fixed by an electrostatic chuck. $C_4F_8$, $CHF_3$, and argon (Ar) gases were introduced into the chamber for the etching. Oxygen ($O_2$) gas was used for plasma ashing, and helium (He) gas was used to cool the wafer. The variable valve automatically controlled the pressure inside the etching chamber during DRIE. This means the valve kept the etching pressure constant. The maximum antenna power was 1 kW, and the maximum bias power was 0.5 kW. In the experiment, the bias and antenna power were fixed at 400 and 600 W, respectively. In addition, the substrate temperature during the DRIE was kept as low as $-20^o C$.

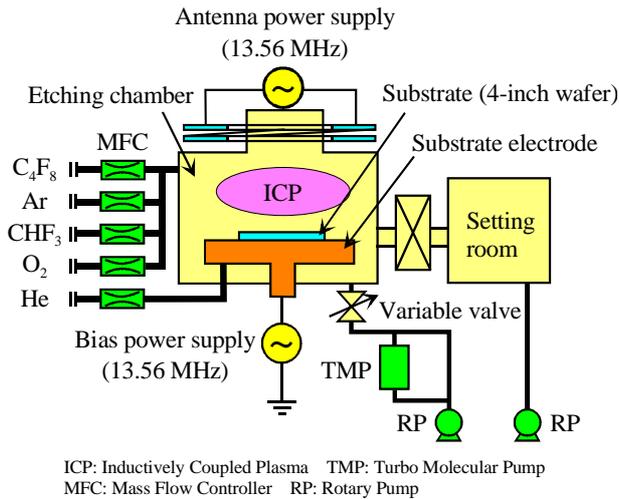

*Figure 1: Dry-etching equipment for DRIE of borosilicate glass (type: NE500, made by ULVAC, Inc.).*

*Table 1: Conditions for DRIE of borosilicate glass.*

| Etching gas | $C_4F_8$/Ar/$CHF_3$ |
| --- | --- |
| Gas pressure | 0.25-0.8 Pa |
| Gas-flow rate | 5-40 sccm |
| Antenna power | 600 W |
| Bias power | 400 W |
| Substrate temperature | -20 degrees Celsius |

### 3. DRIE OF BOROSILICATE GLASS

Figure 2 illustrates the fabrication process for DRIE of borosilicate glass. The process involved using a 200-μm-thick silicon wafer as an etching mask. The silicon wafer was anodically bonded to a borosilicate-glass wafer. In step (a), a 200-μm-deep trench groove was formed on a 300-μm-thick (100)-oriented silicon wafer. Aluminum and silicon-dioxide films were used as an etching mask for DRIE of silicon. In step (b), the DRIE mask was removed, and the wafer was thermally oxidized. Subsequently, a silicon-dioxide film on the back surface was removed. In step (c), a through-hole was formed by etching the wafer with a TMAH solution. In step (d), the silicon wafer was anodically bonded to a borosilicate-glass wafer in this atmosphere. A Pyrex® glass wafer was used as borosilicate glass. The bonding was carried out at 300°C and 300 V in air. In step (e), DRIE of borosilicate glass was implemented using a bonded silicon mask. In step (f), the silicon mask was removed by a KOH solution.

### 4. PROCESS FOR CONTROL OF GROOVE PROFILE

We implemented two fabrication processes to control the groove profile formed using DRIE. One was DRIE with argon added to carbon-fluoride etching gases, i.e., $C_4F_8$ and $CHF_3$ gases; the other was a novel combined process in which DRIE with a mixture of $C_4F_8$ and argon gases and subsequent ultrasonic cleaning in DI water were alternately carried out. The former process means that the etching-gas composition was changed, compared with the previous conditions [7]. The latter process was carried out in step (e), shown in Figure 2. The wafer was cleaned after DRIE of a fixed time. This means that DRIE was again carried out after the ultrasonic cleaning until the depth of the groove reached approximately 300 μm.

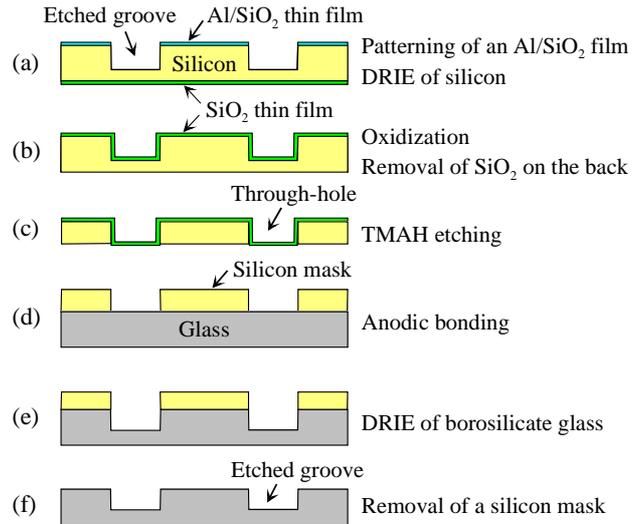

*Figure 2: Fabrication process for DRIE of borosilicate glass.*

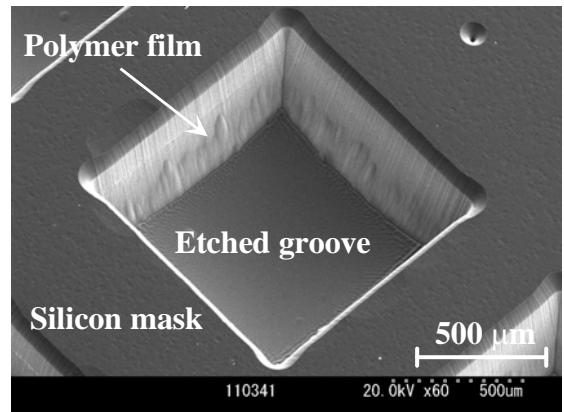

*Figure 3: SEM image of borosilicate-glass groove etched by $C_4F_8$ plasma (opening size: 1mm square; depth: 300 μm).*

### 5. RESULTS AND DISCUSSION

Figure 3 shows an SEM image of a groove etched by $C_4F_8$ plasma before removal of the polymer film and silicon mask. A thick polymer film was observed on the sidewalls of the groove. The typical cross-sectional profile after removal of the silicon mask is shown in

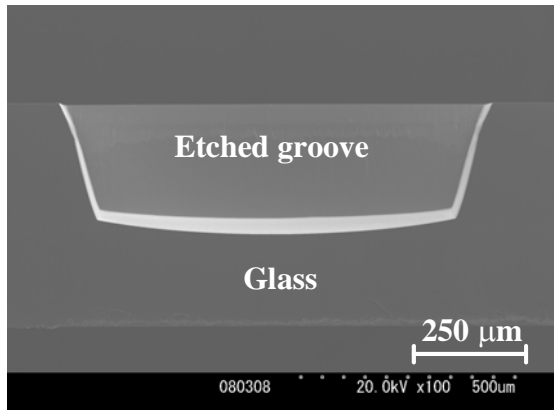

*Figure 4: SEM image of borosilicate-glass groove etched by $C_4F_8$ plasma after removal of a silicon mask by KOH etching (opening size: 1mm square; depth: 300 μm).*

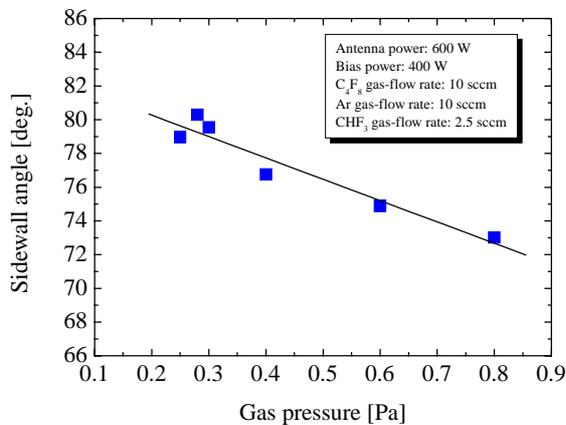

*Figure 5: Measured sidewall angle depending on gas pressure (antenna/bias power: 600/400 W; etching gas: mixture of $C_4F_8$, Ar, and $CHF_3$ gases).*

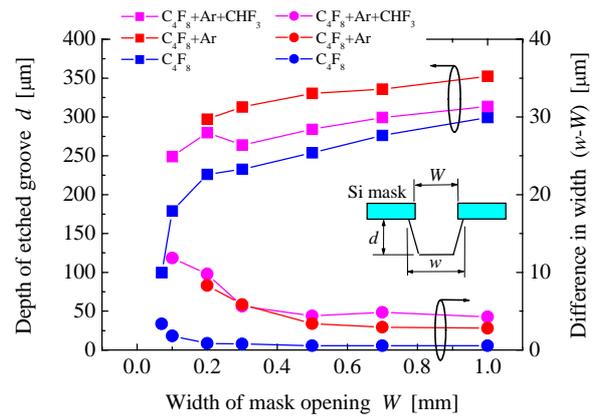

*Figure 6: Profiles of borosilicate-glass grooves etched by $C_4F_8$ plasma, $C_4F_8$+Ar+$CHF_3$ plasma, and combined process ($C_4F_8$ conditions: antenna/bias power: 600/400 W; gas pressure: 0.3 Pa; gas-flow rate: 15 sccm; $C_4F_8$+Ar+$CHF_3$ conditions: antenna/bias power: 600/400 W; gas pressure: 0.28 Pa; gas-flow rate of $C_4F_8$, Ar, $CHF_3$: 10 sccm, 10 sccm, 2.5 sccm; DRIE conditions in combined process: antenna/bias power: 600/400 W; gas pressure: 0.25Pa; gas-flow rate of $C_4F_8$ and Ar: 10 sccm, 10 sccm).*

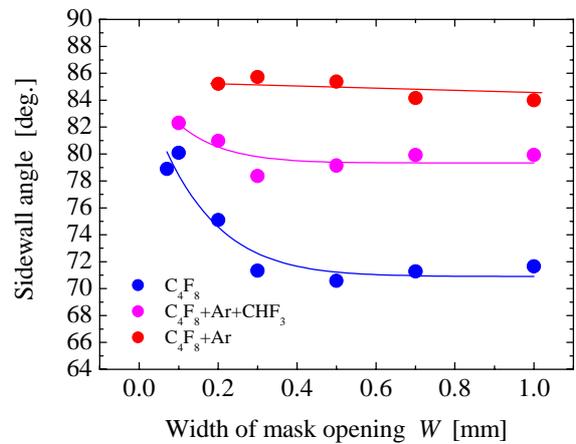

*Figure 7: Measured sidewall angle of borosilicate-glass grooves etched by $C_4F_8$ plasma, $C_4F_8$+Ar+$CHF_3$ plasma, and combined process composed of DRIE with $C_4F_8$+Ar plasma and ultrasonic cleaning in DI water as a function of width of mask opening.*

Figure 4 [7]. The sidewall angle of the groove did not reach more than 80°. In this case, it was around 70° due to the excessive protection of the sidewalls by the polymer film produced by the $C_4F_8$ plasma during DRIE.

Figure 5 shows the dependence of the measured sidewall angle on gas pressure. DRIE was carried out with etching gas composed of $CHF_3$, $C_4F_8$, and argon (Ar). In this case, the gas-flow rate of $CHF_3$, $C_4F_8$, and Ar was 2.5, 10, and 10 sccm, respectively. The figure shows that the necessary condition for obtaining the vertical sidewall is low gas pressure. Accordingly, we need to keep the pressure as low as possible during DRIE to increase the sidewall angle.

Figure 6 shows the measured groove profile representing depth and difference in width between a mask opening and an etched groove. The results shown as $C_4F_8$+Ar+$CHF_3$ mean that DRIE was carried out with argon (Ar) gas added to carbon-fluoride etching gas of $C_4F_8$ and $CHF_3$. The results of $C_4F_8$+Ar mean that the combined process was carried out. In the process, the etching gas of DRIE was composed of both $C_4F_8$ and Ar. The results obtained using DRIE with $C_4F_8$ gas were also plotted for the sake of comparison.

The figure shows that the depth of the etched groove depends on the width of the mask opening. In short, the etching rate gradually increases independent of the com-position of the etching gas as the width increases. This phenomenon indicates aspect-ratio-dependent etching (ARDE) or RIE lag, which can be observed with DRIE of silicon. Concerning the increase in width, mixing argon with the carbon-fluoride gases caused a fairly large difference between the mask opening and the etched groove, compared with the results of $C_4F_8$ gas. This means that the added argon gas contributed to physical etching and that lateral etching slightly advanced. However, the difference in width was kept to less than 12 µm when the depth was around 300 µm. If the depth is less than 300 µm, the difference in width will be much smaller than 12 µm.

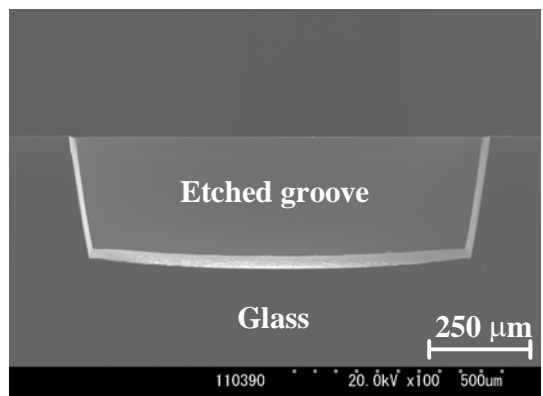

*Figure 8: SEM image of borosilicate-glass groove etched by mixed gas composed of $C_4F_8$, Ar, and $CHF_3$ gases after removal of silicon mask by KOH etching (opening size: 1mm square; depth: 300 µm).*

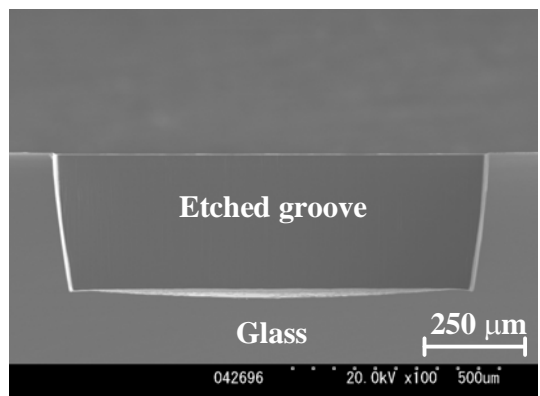

*Figure 9: SEM image of borosilicate-glass groove fabricated by combined process after removal of silicon mask by KOH etching (opening size: 1 mm square; depth: 300 µm).*

Figure 7 shows the dependence of the measured sidewall angle on the width of the mask opening. With $C_4F_8$ gas, the sidewall angle gradually increased and reached its peak at 80° as the width decreased, but with the mixed gases composed of $C_4F_8$, Ar, and $CHF_3$, the obtained angle was approximately 80°, independent of the width of the mask opening. When the width of the mask opening was 1.0 mm, the sidewall angle increased by approximately 8.0°. Accordingly, mixing argon with carbon-fluoride etching gases can improve the sidewall angle. In other words, argon gas needs to be added to control the groove profile. In addition, the combined process resulted in a sidewall angle of 85°, independent of the width. This means the combined process leads to an outstanding improvement in the sidewall angle. To summarize, our fabrication processes can improve the profile and control by removing excessive polymer film on the sidewall.

Figure 8 shows a cross-sectional view of a groove etched with a mixture of $C_4F_8$, Ar, and $CHF_3$ gases. In addition, Figure 9 shows a cross-sectional view of a groove formed using the combined process. The sidewalls in Figures 8 and 9 were slightly inclined off the vertical wall, but the sidewall angle obviously improved. In short, the sidewall angle shown in Figures 8 and 9 is larger than that in Figure 4. Our fabrication processes effectively contribute to removing excessive polymer film on the sidewall. Consequently, we succeeded in forming a 300-µm-deep groove with a sidewall angle of 85° using our process.

## 6. CONCLUSION

Profile control of a borosilicate-glass groove formed by deep reactive ion etching (DRIE) was successfully carried out by effectively removing excessive polymer film produced during DRIE. DRIE with argon added to carbon-fluoride etching gases and a combined process composed of DRIE and ultrasonic cleaning in DI water are effective ways of removing the polymer film. The processes can fabricate a groove with a maximum sidewall angle of 85° independent of its opening width.